%% file: main.tex
\documentclass[runningheads]{llncs}
\usepackage[T1]{fontenc}
\input{preamble}
\usepackage{graphicx}
\usepackage{threeparttable}
\usepackage{booktabs}

\begin{document}
\title{\method}
\author{
Junqi Liu\thanks{Equal contribution. Authors are permitted to list their name first in their CVs.} \and
Dongli He$^{\star}$ \and
Wenxuan Li \and % wli131@jh.edu
Ningyu Wang \and \\
Alan L. Yuille \and
Zongwei Zhou~\Envelope
}

\authorrunning{J. Liu et al.}
\institute{
John Hopkins University \\
\href{mailto:zzhou82@jh.edu} {\textsc{zzhou82@jh.edu}}
}
\maketitle              % typeset the header of the contribution
\begin{abstract}

In this paper, we present a practical approach to improve anatomical shape accuracy in whole-body medical segmentation. Our analysis shows that a shape-focused toolkit can enhance segmentation performance by over 8\%—without the need for model re-training or fine-tuning. In comparison, modifications to model architecture typically lead to marginal gains of less than 3\% \cite{bassi2024touchstone,isensee2024nnu}. Motivated by this observation, we introduce \method, a flexible and easy-to-integrate toolkit designed to refine anatomical shapes. This work highlights the underappreciated value of shape-based tools and calls attention to their potential impact within the medical segmentation community.

\method\ is available at \href{https://github.com/BodyMaps/ShapeKit}{https://github.com/BodyMaps/ShapeKit}

\end{abstract}
\section{Introduction}\label{sec:introduction}

Medical image segmentation sits at the heart of many clinical and research workflows because it turns raw images into \emph{measurable} anatomy and pathology~\cite{litjens2017survey,zhou2022interpreting}. Modern deep networks can already outline dozens of organs in a single pass, yet their \emph{raw} masks still carry avoidable defects—tiny speckles, broken boundaries, swapped left‐and‐right labels, or anatomically impossible shapes \cite{bassi2024label,li2024medshapenet,li2024abdomenatlas,qu2023annotating}. These irregularities lower segmentation accuracy, distort volumetric measurements, and may trigger erroneous downstream decisions~\cite{tajbakhsh2020embracing,ma2021abdomenct,li2024well}.  

Most research tackles these errors by redesigning network architectures or fine‑tuning models \cite{isensee2021nnu,liu2023clip,liu2024universal,zhang2023continual}.  Unfortunately, such changes often deliver only marginal gains (typically \textbf{<3\%} DSC scores if all learning parameters are sufficiently optimized)~\cite{bassi2024touchstone,isensee2024nnu}.  In contrast, a much simpler lever anatomical-aware post‑processing remains under‑explored. We show that correcting masks \emph{after} inference can lift multi–organ DSC score by \textbf{>8\%} without touching the network weights.  This striking gap motivates a fresh look at post‑processing as a first‑class citizen in the segmentation pipeline.

We introduce \method, an easy‑to‑plug‑in toolkit that fixes common shape mistakes in whole‑body CT masks. \method\ offers:  

\begin{itemize}
    \item \textbf{A taxonomy of shape errors} (artifacts, false positives, fragmented structures, redundant lobes, laterality swaps) distilled from large‑scale studies and reader feedback, guiding systematic correction rather than ad‑hoc rules.  
    \item \textbf{General and organ‑specific fixes} built from fast morphological and topological operators that harmonize labels, reconnect broken parts, delete spurious voxels, and re‑assert anatomical plausibility.  
    \item \textbf{Scalability out of the box}: parallel batch processing, vectorized NumPy/SciPy kernels, and minimal memory overhead enable thousands of 3‑D volumes to be cleaned in minutes on a workstation.  
\end{itemize}

By operating solely on the predicted masks, \method\ is \emph{model‐agnostic}. It can upgrade any public or proprietary segmentation models, old or new, with zero re‑training cost—an attractive property for clinical deployment where model updates are tightly regulated.  More broadly, our results rekindle interest in shape reasoning, echoing earlier work on rule‑based and morphological clean‑ups~\cite{gibson2018automatic,reinke2021common} while extending them into a unified, reproducible framework that scales to heterogeneous multi‑organ datasets.

\begin{enumerate}
    \item We quantify how much pure anatomical-aware post‑processing can surpass architecture tweaks, delivering >10\% DSC improvement in widely-adopted, highly-competitive segmentation benchmark.  
    \item We release \method, an open‑source, highly configurable toolkit that standardizes and refines whole‑body CT masks at scale.  
    \item We provide an extensive error taxonomy and ablation study to guide future work on shape‑aware medical segmentation.  
\end{enumerate}

\section{Related Work}\label{sec:related_work}

Post-processing is a critical but often under-emphasized step in medical image segmentation. Early techniques relied heavily on manual correction and handcrafted rules to eliminate artifacts and fix mislabeled regions \cite{ronneberger2015u}. With the rise of deep learning, most effort shifted toward improving segmentation networks \cite{zhang2024leveraging,zhou2019unet++}, yet the final predictions from even the strongest models frequently suffer from shape errors—such as disconnected fragments, implausible geometry, or label confusion---that require correction downstream \cite{ma2021abdomenct}.

Recent toolkits, such as nnU-Net \cite{isensee2021nnu}, MONAI \cite{cardoso2022monai}, and DAP-Atlas \cite{gatidis2022whole}, provide basic post-processing utilities, including connected component filtering and label remapping. However, these modules are often fixed in design and lack anatomical reasoning. More advanced efforts have explored shape priors \cite{heinrich2016deformable} or statistical harmonization \cite{wang2022cross,gibson2018automatic}, but these are usually domain-specific or computationally intensive. Despite growing datasets and deployment in clinical workflows, few systems offer a general, extensible, and anatomically aware post-processing pipeline capable of handling diverse whole-body CT data.

\method\ addresses this gap. Unlike prior toolkits, it introduces a detailed taxonomy of shape errors, a set of modular correction functions, and both general-purpose and organ-specific refinements—all with scalability and ease-of-use in mind. This toolkit is designed not only to clean segmentation outputs but also to enforce anatomical plausibility at scale.

\section{\method}\label{sec:method}

\method\ is a lightweight, modular toolkit for anatomical shape correction in CT segmentation. It offers two tiers of functions: general-purpose operations for common cleanup tasks and organ-specific refinements for anatomical accuracy. The entire system is plug-and-play, requiring no retraining or model changes.

\begin{figure}[t]
    \centering
    \includegraphics[width=\linewidth]{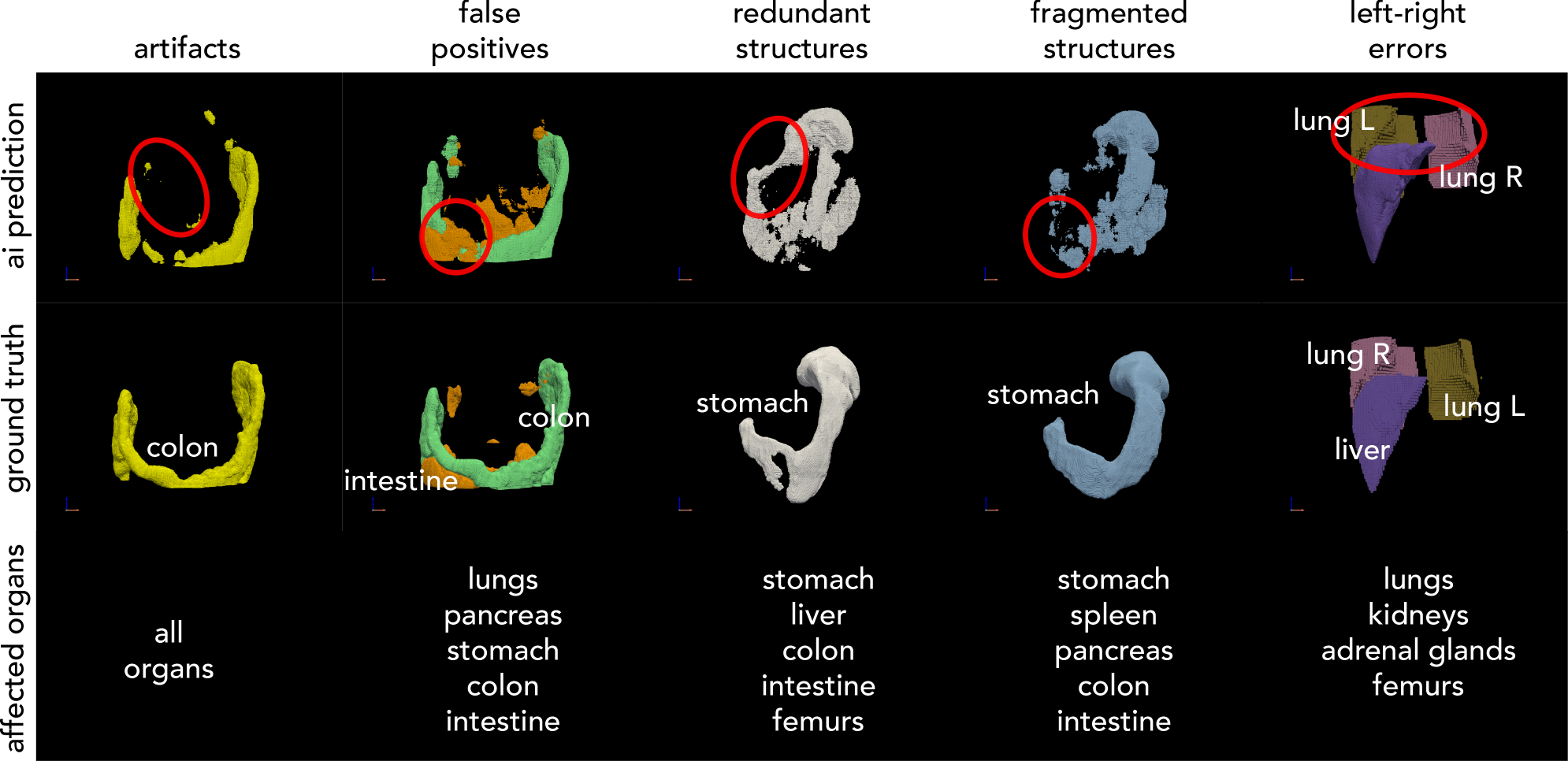}
    \caption{Types and distribution of common segmentation errors identified across different abdominal organs in the VISTA3D dataset. Categories include false positives, redundant or fragmented structures, and left-right errors. Organs are grouped by frequency of impact}
    \label{fig:errors}
\end{figure}

\subsection{A Taxonomy of Shape Errors}

Segmentation outputs often contain structural inconsistencies that deviate from expected anatomical patterns. These shape errors are not just cosmetic—they can distort volumetric measurements, hinder morphometric analysis, and mislead downstream algorithms. Based on analysis of large-scale whole-body CT datasets, including expert feedback and manual inspection, we categorize the most frequent error modes into five types, visualized in Fig.~\ref{fig:errors}:

\begin{enumerate}
    \item \textbf{Artifacts:} These appear as small, isolated voxel clusters that float outside the anatomical region of interest. For example, speckle-like debris near the liver or stomach (Fig. 1, column 1) often arise due to imaging noise or minor model misfires. While small, these anomalies can falsely inflate organ volumes and confuse spatial reasoning tasks.
    \item \textbf{False Positives:} These are anatomically implausible regions labeled as an organ, but which actually belong to unrelated tissue. As shown in Fig. 1 (column 2), a portion of the intestines is mistakenly labeled as colon, or parts of the stomach leak into surrounding space. These misclassifications are often more prominent in low-contrast areas or near organ boundaries.
    \item \textbf{Redundant Structures:} These errors extend the predicted shape into abnormal or biologically impossible regions. For instance, the segmentation may include duplicated lobes of the lung, or extensions of the colon that protrude into the abdominal cavity without anatomical basis (Fig. 1, column 3). Such predictions violate known organ morphology and must be removed or corrected.
    \item \textbf{Fragmented Structures:} Rather than being detected as a single continuous object, the organ is split into disconnected fragments. Figure 1 (column 4) shows this clearly in the pancreas and colon—organs that are anatomically continuous but predicted as multiple unconnected blobs. This affects both the anatomical integrity and the reliability of region-based statistics.
    \item \textbf{Left-Right Errors:} Perhaps the most dangerous type of mistake in clinical contexts, these occur when left/right anatomical labels are swapped or missing entirely. In Figure 1 (column 5), the left and right lungs are either merged or mislabeled, and the liver appears on the wrong side. Such spatial inversion could severely impact tasks like surgical planning or disease localization.
\end{enumerate}
    
By systematically identifying these categories, \method\ enables a structured and reproducible approach to post-processing, moving beyond ad-hoc filtering or patch-based fixes. Each of the functions is designed to directly address one or more of these common shape inconsistencies.

\subsection{General \method\ Functions}

\method\ provides a suite of general-purpose functions designed to correct common structural errors in organ segmentation masks. These functions are model-agnostic and operate directly on binary or multi-label masks produced by any segmentation algorithm. By focusing on structural integrity and anatomical plausibility, they help convert raw predictions into clinically usable outputs. Each function is modular and can be flexibly composed with others, enabling users to tailor the pipeline to different datasets or application needs.

\begin{enumerate}
    \item \textbf{Removal of Small Components.} One of the most frequent segmentation artifacts involves the presence of small, disconnected regions that do not belong to the actual organ. These tiny speckles or blobs are often caused by noise in the image or minor classification errors by the model. \method\ removes such components based on a user-defined volume threshold. This ensures that only sufficiently large, anatomically meaningful regions are retained, thereby improving the accuracy and cleanliness of the final mask.
    \item \textbf{Suppression of Non-Dominant Components.} In cases where multiple disconnected regions of varying sizes are labeled as the same organ, \method\ retains only the most dominant ones. This function ranks connected components by size and preserves only the largest (e.g., top 1 or top 2), assuming that smaller components are likely to be spurious. This is particularly useful for organs such as the liver or pancreas, which should appear as single contiguous structures.
    \item \textbf{False Positive Reassignment.} False positives occur when a region is mistakenly labeled as an organ but actually corresponds to nearby tissue. To address this, \method\ analyzes spatial relationships between predicted regions and known anatomical adjacency. If a component appears too far from the expected location of a given organ or is closer to another organ, it is relabeled accordingly. This function improves specificity and reduces label leakage between adjacent organs, which is especially important in complex regions like the abdomen.
    \item \textbf{Left–Right Separation and Validation.} For symmetric organs—such as lungs, kidneys, and adrenal glands—laterality matters. \method\ introduces a spatially aware function to divide merged predictions into left and right components. It first splits the mask along a reference anatomical axis (typically the sagittal plane), then assigns each component to the appropriate side based on spatial proximity to a reliable landmark, such as the liver (assumed to reside on the right side of the body). This prevents left-right confusion and ensures consistency in downstream analysis or reporting.
    \item \textbf{Merging Fragmented Structures.} Sometimes, an organ is predicted as multiple broken parts that should, in reality, be a single entity. \method\ detects when multiple disconnected components of the same organ are in close spatial proximity and merges them into a unified structure. This enhances continuity and avoids misleading interpretations of organ shape, size, or completeness.
    % \item \textbf{Filling Internal Voids.} Segmentation outputs may occasionally include hollow regions inside an organ mask—small cavities that disrupt the expected solid structure. These voids can arise from thresholding errors or imaging inconsistencies. \method\ includes a function to fill such holes, ensuring that organs like the spleen or bladder are represented as continuous, enclosed volumes.
\end{enumerate}

Together, these general functions address the majority of shape errors that plague raw segmentation masks. They require no training or annotation effort and can be applied to any organ or dataset. 
% Their plug-and-play design allows for seamless integration into existing pipelines, offering users a principled way to enforce anatomical consistency at scale.

\subsection{Specialized \method\ Functions}

While general post-processing functions help fix common shape issues across all organs, certain errors are unique to specific anatomical structures. For example, the lungs require careful left–right separation, the intestines are prone to fragmentation, and the pancreas may appear as several disconnected parts due to its irregular shape. To address these challenges, \method\ offers specialized functions tailored to the geometry and spatial context of individual organs.

These specialized functions build upon the general tools but include additional logic based on organ-specific knowledge. For example, the toolkit recognizes that the liver is always located on the right side of the body. This anatomical fact is used to help resolve left-right confusion in neighboring organs, such as the lungs or kidneys. A few key examples of these specialized functions include:

\begin{enumerate}
    \item \textbf{Left–Right Assignment Using the Liver.} \method\ first divides symmetric organs (like the lungs or kidneys) into left and right sides by looking at their location in the scan. But because some patients are positioned differently, this step may not always be accurate. To make it more reliable, \method\ compares the position of these organs to the liver. Since the liver is known to sit on the right side of the body, it can act as a reference point to confirm or correct the organ labels. This is especially helpful when both sides of an organ are merged or mislabeled.
    % \item \textbf{False Positive Reassignment for Neighboring Organs.} Some organs, like the pancreas, colon, and duodenum, are located very close to each other and may be mistakenly labeled as one another. \method\ uses a proximity-based method to reassign mislabeled regions. For example, if a region labeled as “pancreas” is actually closer to the duodenum, it will be relabeled accordingly. This spatial logic helps reduce label confusion in crowded abdominal areas.
    % \item \textbf{Fragment Merging for Elongated Organs.} Organs such as the pancreas or intestines are long and thin, often leading to broken predictions. \method\ scans for disconnected parts of the same organ and merges them into a single connected structure when they are close enough together. This results in more realistic segmentations that match the true anatomy.
    % \item \textbf{Removal of Implausible Additions.} Sometimes, a segmentation mask contains extra "lobes" or projections that do not belong to any real organ. These redundant structures are especially common in the stomach or liver regions. \method\ checks for such protrusions and removes them if they extend beyond plausible anatomical boundaries. This helps ensure that the organ shape looks reasonable.
    \item \textbf{Retaining Key Structures While Filtering Noise.} For each organ, \method\ includes size and shape thresholds to decide what to keep and what to discard. For instance, when cleaning up the lungs or kidneys, the toolkit keeps only the two largest components (left and right) and removes smaller, likely incorrect pieces. These rules are informed by the typical size and count of organs and help avoid over-cleaning or under-filtering.
\end{enumerate}

Importantly, all these specialized corrections are modular. Users can choose which organs to process and which rules to apply using a configuration file. This flexible setup makes it easy to apply \method\ to new datasets or specific clinical tasks without re-writing the code. The modular design also allows users to build custom pipelines by combining only the needed functions, similar to assembling blocks in a Lego set.

In summary, the organ-specific functions bring domain knowledge into the correction process. By understanding how each organ should appear and where it should be located, \method\ provides precise and trustworthy refinements that go beyond generic image processing.

\subsection{Computational Optimization}

To ensure scalability across large datasets, \method\ incorporates several key computational optimizations.

First, the pipeline supports parallel execution using Python’s multiprocessing, enabling batch processing of thousands of CT volumes. Each scan is handled independently, allowing efficient workload distribution across CPU cores—critical for time-sensitive clinical settings or large-scale studies.

Second, \method\ relies on vectorized operations via high-performance libraries such as NumPy and SciPy. These implementations eliminate slow Python loops and leverage optimized C-backed routines for morphological operations, connected component analysis, and spatial filtering.

Third, to reduce memory usage, \method\ processes only the relevant masks in compressed data types (e.g., bool, uint8) and clears intermediate arrays after use. The system also employs lazy evaluation: if no correction is needed for a particular function, the step is skipped automatically.

Finally, the modular design allows users to enable only selected functions via a configuration file, avoiding unnecessary computation and making the pipeline efficient and adaptable to diverse datasets.

\section{Experiment \& Result}\label{sec:experiment}

\noindent\textbf{Dataset.} We used a subset of the JHH dataset~\cite{park2020annotated,xia2022felix} (denoted as Dataset A) and a subset of AbdomenAtlasPro~\cite{li2024abdomenatlas,bassi2025radgpt,li2025scalemai} (denoted as Dataset B) to evaluate the \method\ pipeline. The JHH dataset includes 5,160 contrast-enhanced CT scans (venous and arterial phases) from international sites, centralized at Johns Hopkins Hospital. Each scan contains metadata such as age, race, gender, and diagnosis. Voxel-wise annotations were manually created by radiologists and verified by senior experts. All scans were de-identified and approved for use by the Johns Hopkins IRB (IRB00403268). AbdomenAtlasPro comprises 20,460 abdominal CT volumes from 112 hospitals across diverse populations. It includes 673K segmentation masks across 22 anatomical structures, generated through a two-stage pipeline combining expert annotation and AI-assisted refinement. And all data were anonymized prior to public release. For the purpose of evaluation, representative subsets from both the JHH and AbdomenAtlasPro datasets were selected and exclusively reserved for testing.

\smallskip\noindent\textbf{Evaluation Metric.} To systematically evaluate the effectiveness of the \method\ pipeline, we employed the Dice–Sørensen coefficient (DSC) \cite{dice1945measures} as the primary performance metric. For each subject, DSC values were computed organ-wise, both before and after applying the correction functions. This comparative analysis enabled a detailed quantification of segmentation improvements attributable to shape corrections, capturing both global and localized refinements. By assessing changes in DSC across organs, we were able to identify which anatomical structures benefited most from specific correction strategies and to characterize the overall robustness of the pipeline.

\begin{table}[t]
    \centering
    \scriptsize
    \caption{
    Comparison of DSC scores before and after applying \method\ for multiple organs.
    }
    \begin{tabular}{p{0.22\linewidth}P{0.12\linewidth}P{0.12\linewidth}P{0.12\linewidth}P{0.12\linewidth}P{0.12\linewidth}P{0.12\linewidth}}
    \toprule
    \textbf{Dataset A} & gall bladder & pancreas & lung L & stomach & lung R & duodenum\\
    \midrule
    VISTA3D \cite{he2024vista3d} & 71.7 & 79.3 & 87.5  & 91.4 & 89.0 & 72.9\\
    VISTA3D+\method\ & 79.7 & 84.5 & 91.0  & 93.7 & 91.2  & 75.0 \\
    $\Delta$ & \textbf{\textcolor{blue}{+8.0}} & \textbf{\textcolor{blue}{+5.2}} & \textbf{\textcolor{blue}{+3.5}} & \textbf{\textcolor{blue}{+2.3}} & \textbf{\textcolor{blue}{+2.2}}  & \textbf{\textcolor{blue}{+2.1}} \\
    % \bottomrule
    \midrule
    \textbf{Dataset B} & gall bladder & pancreas & colon & lung R & lung L & stomach \\
    \midrule
    VISTA3D \cite{he2024vista3d} & 68.1 & 73.5 & 80.1 & 88.6 & 87.8 & 91.3\\ 
    VISTA3D+\method & 76.9 & 79.0 & 84.2 & 91.7 & 90.5 & 93.6\\
    $\Delta$ & \textbf{\textcolor{blue}{+8.8}} & \textbf{\textcolor{blue}{+5.5}} & \textbf{\textcolor{blue}{+4.1}} & \textbf{\textcolor{blue}{+3.1}} & \textbf{\textcolor{blue}{+2.7}}  & \textbf{\textcolor{blue}{+2.3}} \\
    \bottomrule
    \end{tabular}
    \label{tab:reader_studies}
\end{table}

\subsection{Quantitative Results} 

Table~\ref{tab:reader_studies} summarizes the Dice–Sørensen Coefficient (DSC) scores across multiple organs before and after applying \method\ to segmentation outputs from the VISTA3D model~\cite{he2024vista3d}. The evaluation was conducted on two independent test sets, i.e., Dataset A and Dataset B, each sampled from different clinical sources and imaging protocols. Importantly, both datasets are out-of-distribution (OOD) relative to the training data used by VISTA3D, making this a strong test of generalization.

Across both datasets, \method\ consistently improves segmentation accuracy, reinforcing its effectiveness as a post-processing layer under distribution shift. In Dataset A, the most notable gains are observed in the gall bladder (+8.0 DSC) and pancreas (+5.2 DSC), organs known to pose challenges due to their small size and variable appearance. Improvements in lung L/R, stomach, and duodenum (ranging from +2.1 to +3.5) suggest that \method\ also successfully corrects subtle boundary errors and enforces anatomical symmetry.

In Dataset B, similar trends are observed, with even higher gains in gall bladder (+8.8) and pancreas (+5.5), as well as in more complex, elongated structures like the colon (+4.1). These results indicate that \method\ is not only effective in refining predictions from in-distribution data, but also robust to changes in data domain, scanner protocol, or patient population.

One particularly noteworthy observation is that even high-performing organs—such as the lungs and stomach, which already have DSCs exceeding 85\%, experience meaningful gains. This suggests that \method\ is not simply compensating for weak segmentations, but systematically improving mask plausibility across the board by reducing noise, correcting structural errors, and restoring anatomical integrity.

\begin{figure}[htbp]
    \centering
    \includegraphics[width=\linewidth]{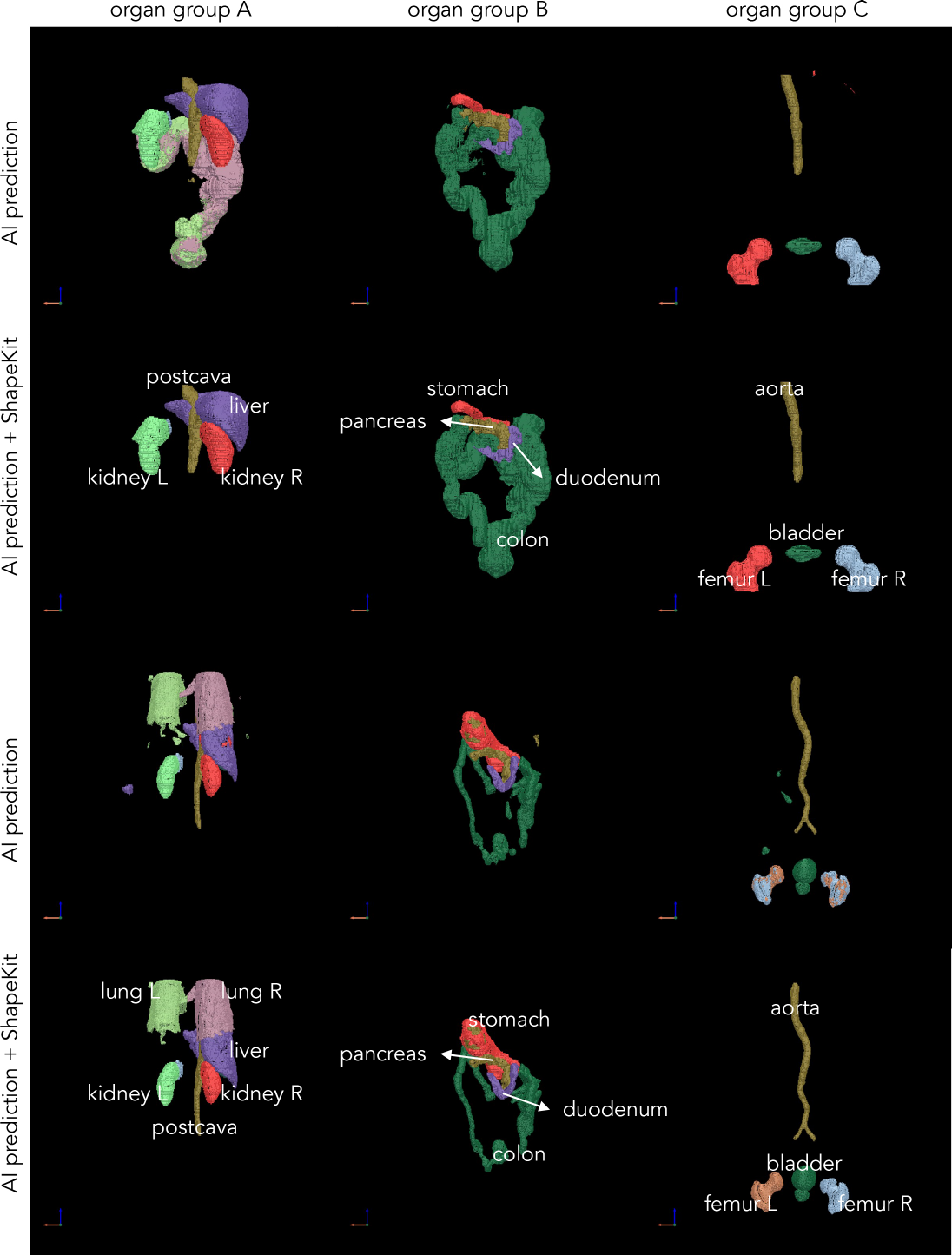}
    \caption{Visualization of two representative cases segmented by ai prediction, with and without \method. To facilitate comparison, annotated organs are grouped into three categories: Group A (lungs, esophagus, liver, gall bladder, hepatic vessel, portal and splenic veins, kidneys, adrenal glands, postcava), Group B (stomach, pancreas, duodenum, colon, intestine, rectum), and Group C (femurs, aorta, celiac trunk, bladder, prostate). In the post-processed segmentations, organ labels are overlaid for clarity.}
    \label{fig:visualization}
\end{figure}

\subsection{Visualization Results} 

Figure~\ref{fig:visualization} illustrates qualitative comparisons of segmentation results before and after applying \method. These examples highlight common failure modes—such as artifacts, false positives, fragmented regions, and left-right label errors—across a variety of organs including the lungs, pancreas, colon, and stomach.

In the raw predictions, we observe noisy speckles around the liver and intestines, discontinuous pancreas masks, and incorrect left-right labeling in symmetric organs like the lungs and kidneys. Such errors degrade anatomical coherence and can mislead downstream measurements, especially for shape-sensitive tasks like volume estimation or surgical planning.

After post-processing with \method, these issues are significantly mitigated. Fragmented organs are reconnected, redundant or implausible structures are removed, and left/right sides are correctly resolved based on liver positioning. Notably, organs such as the pancreas and colon—which are particularly prone to false positives and fragmentation—show visibly more compact and anatomically realistic segmentations. Improvements in high-performing organs like the lungs and stomach further demonstrate that \method\ refines even strong predictions by enforcing symmetry and boundary regularity.

These visual results confirm that \method\ not only removes superficial noise but meaningfully improves anatomical plausibility, contributing to more reliable and interpretable outputs for clinical and quantitative analysis.

\section{Discussion \& Conclusion}\label{sec:discussion_conclusion}

We introduced \method, a flexible, model-agnostic toolkit that systematically corrects shape-related errors in whole-body CT segmentation. Through a suite of modular correction functions, \method\ improves anatomical plausibility without requiring model retraining or architectural changes. Our results show consistent DSC improvements in two out-of-distribution datasets, highlighting its robustness and ease of integration into existing segmentation models.

A key insight from this work is that post-processing, often overlooked, can deliver meaningful improvements—especially in challenging or fragmented predictions. \method\ addresses not only cosmetic issues like speckles or label leakage but also enforces deeper structural consistency, such as left-right correctness and organ continuity.

Importantly, anatomically accurate shape masks are increasingly essential beyond evaluation. Recent efforts in synthetic data generation, such as NVIDIA’s MAISI platform \cite{guo2025text2ct,guo2025maisi}, and other related studies \cite{mao2025medsegfactory,chen2024towards,hu2023label,lai2024pixel,yang2025medical,li2024text,du2024boosting}, rely on precise segmentation shapes to synthesize realistic CT volumes. In these pipelines, toolkits like \method\ serve as a critical preconditioning step—ensuring that imperfect model outputs do not propagate into downstream training data. Similarly, recent work by Li~\etal~\cite{li2024medshapenet} shows that shape consistency enhances model robustness and domain transfer. \method\ contributes directly to this by refining outputs in a reproducible, interpretable manner.

Despite its strengths, \method\ is currently rule-based and may oversimplify segmentations in highly complex structures, such as the colon. Future extensions could integrate learned shape priors or uncertainty-aware corrections to further improve generalization across anatomical variations and pathologies.

In conclusion, \method\ elevates the role of post-processing in medical image segmentation. By enforcing anatomical structure at scale, it not only enhances segmentation quality but also supports broader applications such as data curation, synthetic image generation, and shape-aware modeling. We release \method\ as an open-source toolkit and hope it serves as a foundation for future shape-centric advances in medical AI.

\begin{credits}
\subsubsection{\ackname} This work was supported by the Lustgarten Foundation for Pancreatic Cancer Research and the National Institutes of Health (NIH) under Award Number R01EB037669. We would like to thank the Johns Hopkins Research IT team in \href{https://researchit.jhu.edu/}{IT@JH} for their support and infrastructure resources where some of these analyses were conducted; especially \href{https://researchit.jhu.edu/research-hpc/}{DISCOVERY HPC}.

\subsubsection{\discintname}
The authors declare no competing interests.
\end{credits}
% \clearpage
\bibliographystyle{splncs04}
\bibliography{zzhou,refs}

\input{appendix}

\end{document}

%% file: preamble.tex
\usepackage{float}
\usepackage{pifont}
\usepackage{footnote}
\usepackage{enumitem}
\usepackage{bm}
\usepackage{arydshln}
\usepackage{booktabs}
\usepackage{multicol}
\usepackage{multirow}
\usepackage{color}
\usepackage{xcolor}     
\usepackage{colortbl}
\usepackage{soul}
\usepackage{bbding}
\usepackage{makecell}
\usepackage{mathtools}
\usepackage{imakeidx}
\usepackage{amssymb}
\usepackage{graphicx}
\usepackage{amsmath}
\usepackage{threeparttable}
\usepackage{bbding} 
\definecolor{citecolor}{HTML}{0071BC}
\definecolor{linkcolor}{HTML}{ED1C24}
\usepackage[colorlinks,
            anchorcolor=red,
            citecolor=citecolor, 
            linkcolor=linkcolor,
            ]{hyperref}
\makeindex
\usepackage{arydshln}
\usepackage{lipsum}
\usepackage[toc]{multitoc}
\usepackage[edges]{forest}
\usepackage[normalem]{ulem}

\usepackage{bbding}
\usepackage[most]{tcolorbox}

\usepackage{algorithm}
\usepackage{algorithmic}

\usepackage{minitoc}
\usepackage[toc,page,header]{appendix}

\definecolor{orchid}{rgb}{0.85, 0.44, 0.84}
\definecolor{rubinred}{rgb}{0.82, 0.0, 0.28}
\definecolor{flagship}{rgb}{0.93, 0.06, 0.41}
\definecolor{radiologist}{rgb}{0.50, 0.50, 1}

\newcommand{\etal}{\mbox{et al.}}

\definecolor{YT}{HTML}{002FA7}

\newcommand{\method}{{\fontfamily{ppl}\selectfont
ShapeKit}}

\newcolumntype{P}[1]{>{\centering\arraybackslash}p{#1}}
\newlength\savewidth
  

%% file: appendix.tex
\clearpage
\section*{Appendix: Major \method \  Functions}\label{sec:appendix1}

\subsection*{Remove Small Components}

This function removes small disconnected regions from a binary mask based on a specified volume threshold. It is commonly used to eliminate noisy artifacts, improving the anatomical plausibility of the segmentation.

\begin{verbatim}
def remove_small_components(
    mask: np.array, 
    threshold: int):
    """
    Remove small components based on a given volume threshold.

    Args:
        mask (np.array): Binary 3D array.
        threshold (int): Minimum number of voxels a component 
        must have to be kept.
        
    Returns:
        np.array: A cleaned binary mask.
    """
\end{verbatim}

\subsection*{False Positive Reassignment}

This function aims to correct false positives that arise when components are mistakenly assigned to the wrong organ. If a component is found to be closer to a neighboring organ, it is reassigned accordingly. This improves segmentation specificity and reduces cross-organ mislabeling.

\begin{verbatim}
def reassign_false_positives(
    segmentation_dict: dict, 
    organ_adjacency_map: dict, 
    check_size_threshold=500):
    """
    Reassign false positives between anatomically adjacent organs.

    Args:
        segmentation_dict (dict): A dictionary mapping organ 
        names to binary masks. 
        organ_adjacency_map (dict): A dictionary defining 
        spatial adjacency between organs. 
        check_size_threshold (int, optional): Minimum component 
        size to consider for reassignment. 
                                              
    Returns:
        dict: Updated segmentation dictionary.
    """
\end{verbatim}

\subsection*{Suppress Non-dominant Components}

This function retains only the top \texttt{N} largest connected components in a binary mask, removing smaller fragments. It is commonly used to eliminate noise or fragmented structures while preserving the dominant anatomical region.

\begin{verbatim}
def suppress_non_largest_components_binary(
    mask:np.array, 
    keep_top=2):
    """
    Suppress all but the top-N largest connected components
    in a binary mask.
    
    Args:
        mask (np.ndarray): Binary mask
        keep_top (int): Number of largest components to keep.

    Returns:
        np.ndarray: Cleaned binary mask.
    """
\end{verbatim}

\subsection*{Left-right Division Check}

This step separates symmetric organs (e.g. lungs, kidneys) into left and right components based on their spatial position. Components are split along a chosen axis (typically the sagittal plane), and laterality is verified using the anatomical location of a reference organ such as the liver, which is assumed to reside on the right side.

\begin{verbatim}
def split_right_left(mask, AXIS=0):
    """
    Splits a symmetric organ mask into right and left components 
    along a specified axis.

    Each connected component in the mask is assigned to either 
    the left or right side based on the mean coordinate of its 
    voxels along the specified axis.

    Args:
        mask (np.ndarray): Binary 3D mask containing 
        the merged organ(s).
        AXIS (int): Axis along which to perform the split. 
        Default is 0 (left-right).

    Returns:
        right_mask (np.ndarray): Binary mask.
        left_mask (np.ndarray): Binary mask.
    """
\end{verbatim}

\vspace{10pt}
\begin{verbatim}
def reassign_left_right_based_on_liver(
    right_mask: np.array, 
    left_mask: np.array, 
    liver_mask: np.array):
    """
    Reassign left and right masks based on proximity to the liver.
    Liver is assumed to always be on the right side anatomically.

    Args:
        right_mask (np.ndarray): Binary mask for the presumed 
        right-side organ.
        left_mask (np.ndarray): Binary mask for the presumed 
        left-side organ.
        liver_mask (np.ndarray): Binary mask for the liver 
        (used as spatial reference).
        
    Returns:
        corrected_right_mask, corrected_left_mask : np.ndarray
    """
\end{verbatim}